\documentclass[]{spie}  

\usepackage{amsmath,amsfonts,amssymb}
\usepackage{graphicx}
\usepackage[colorlinks=true, allcolors=blue]{hyperref}

\title{Towards new servo control algorithms at the TNG telescope}

\author[a]{P. Schipani}
\author[b]{M. Gonzalez}
\author[a]{F. Perrotta}
\author[a]{S. Savarese}
\author[a]{M. Colapietro}
\author[b]{A. Ghedina}
\author[b]{M. Hernandez Diaz}
\author[b]{H. Ventura}
\affil[a]{INAF - Osservatorio Astronomico di Capodimonte, Salita Moiariello 16, Naples, Italy}
\affil[b]{INAF - FGG, TNG, Rambla J.A. Fernández Pérez 7, Breña Baja (TF), Spain}

\authorinfo{Send correspondence to: pietro.schipani@inaf.it}

\begin{document} 
\maketitle

\begin{abstract}
The servo control algorithms of the TNG, developed in the nineties, have been working for more than 20 years with no major updates. The original hardware was based on a VME-bus based platform running a real time operating system, a rather popular choice for similar applications at the time. Recently, the obsolescence of the hardware and the lack of spares pushed the observatory towards a complete replacement of the electronics that is now being implemented in steps, respecting the basic requirement of never stopping the observatory night operations. Within the framework of this major hardware work, we are taking the opportunity to review and update the existing control schemes. This servo control update, crucial for the telescope performance, envisages a new study from scratch of the controlled plant, including a re-identification of the main axes transfer functions and a re-design of the control filters in the two nested position and speed loops. The ongoing work is described, including preliminary results in the case study of the azimuth axis and our plans for possible further improvements.  
\end{abstract}

\keywords{TNG, Telescope, Tracking, Servo Control}

\section{INTRODUCTION}
\label{sec:intro}  
The 3.5-m TNG telescope saw first light in La Palma (Canary Islands) at Roque de Los Muchachos observatory in 1998. Its axes control system had been realized few years before: all the control hardware had been installed in 1995, completing the procurement of components and custom parts during the preliminary integration in Italy. In the original design, the speed loop control had been implemented adopting a custom hardware developed by the Sierracin Magnedyne company, including motors, power amplifiers and control CPUs, whereas the position loop had been realized in-house by the Italian community, using all commercial components based on the VME bus technology and a real-time operating system: this system has been working every night for more than 20 years. 

Only few relatively minor interventions have been necessary over the years, but after more than two decades, the observatory was running out of spare parts. Soon after the telescope realization, the stock could no longer be updated for the custom parts. Of course, over the years this became a critical issue because of aging and obsolescence, including the commercial components as well. Thus, recently the observatory started a project to replace all the control hardware components but motors and feedback sensors: this work is described in Ref.~\citenum{Gonz20}. Obviously, this kind of modernization efforts are common for telescopes operating since a few decades (e.g. see Ref.~\citenum{Abbott08,Spray16,Krasuski18}).

Within this effort, it was decided to maintain the same software, which had proved to be effective and reliable for many years, so avoiding a full recommissioning of the system. Thus, the control software has been ported to the new platform with a heavy and accurate work of translation from the old real-time textual environment to the National Instruments LabVIEW graphical programming language. Thus, the telescope is still running the “same” axes control software albeit under a totally different environment. 

As the whole intervention involves critical system components (e.g. the axes drives), possible effects on the axes transfer function, and consequently on the servo control performance, can be expected. This motivated the team to plan a complete check and redesign of the servo control algorithms, under the strict requirement of never stopping the telescope operations. This is the main objective of this paper, which describes the case of the azimuth axis.

   \begin{figure} [ht]
   \begin{center}
   \begin{tabular}{c} 
   \includegraphics[height=11cm]{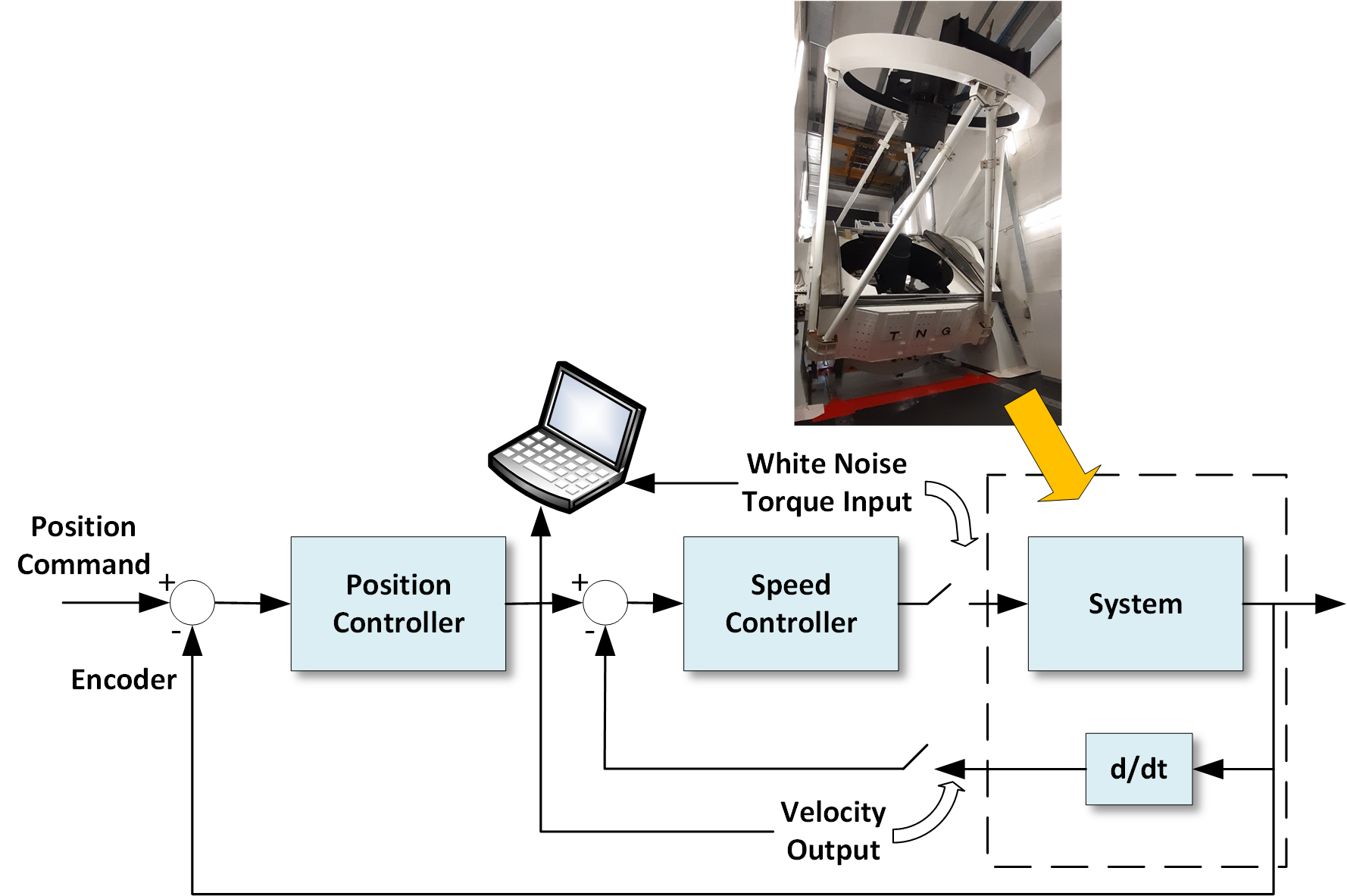}
   \end{tabular}
   \end{center}
   \caption[example] 
   { \label{fig:IdScheme}
Open loop model identification. The white noise is injected into the torque input, the differentiated encoder signal is recorded synchronously as system output. The I/O sample pairs are analyzed in Matlab through a laptop computer. }
     \end{figure}

   \begin{figure} [ht]
   \begin{center}
   \begin{tabular}{c} 
   \includegraphics[height=12cm]{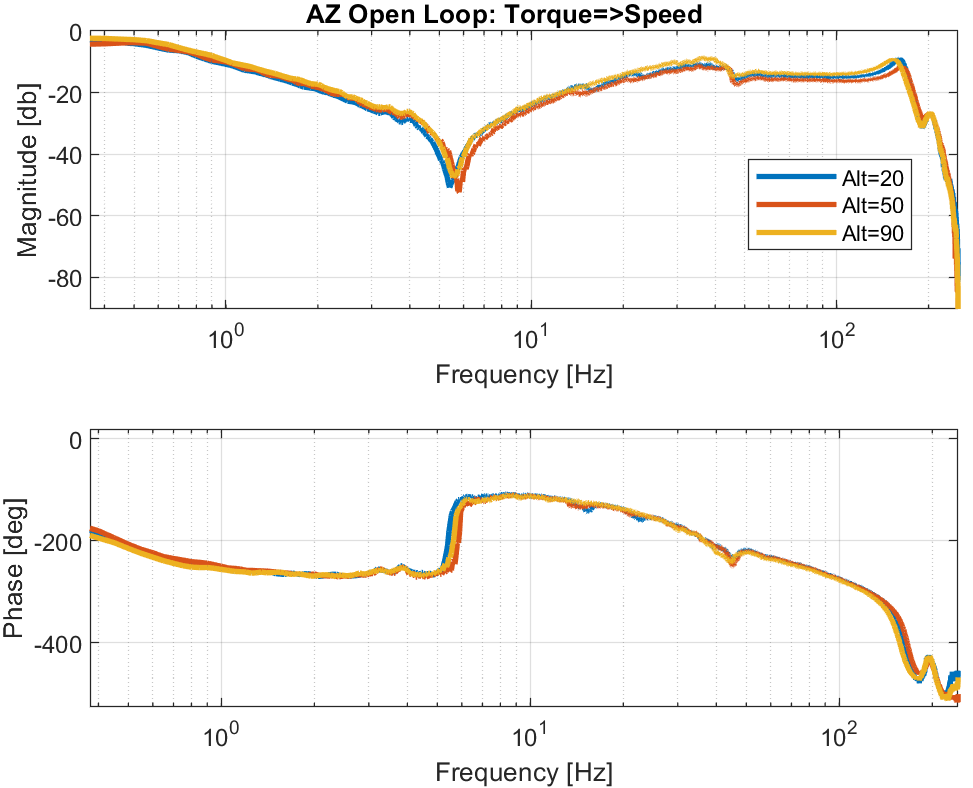}
   \end{tabular}
   \end{center}
   \caption[example] 
   { \label{fig:TF}
Open-loop transfer functions of the azimuth axis at three different altitude angles: ALT=90$^\circ$ (Zenith), ALT=50$^\circ$, ALT=20$^\circ$ (observational limit).}
     \end{figure}

   \begin{figure} [ht]
   \begin{center}
   \begin{tabular}{c} 
   \includegraphics[height=7cm]{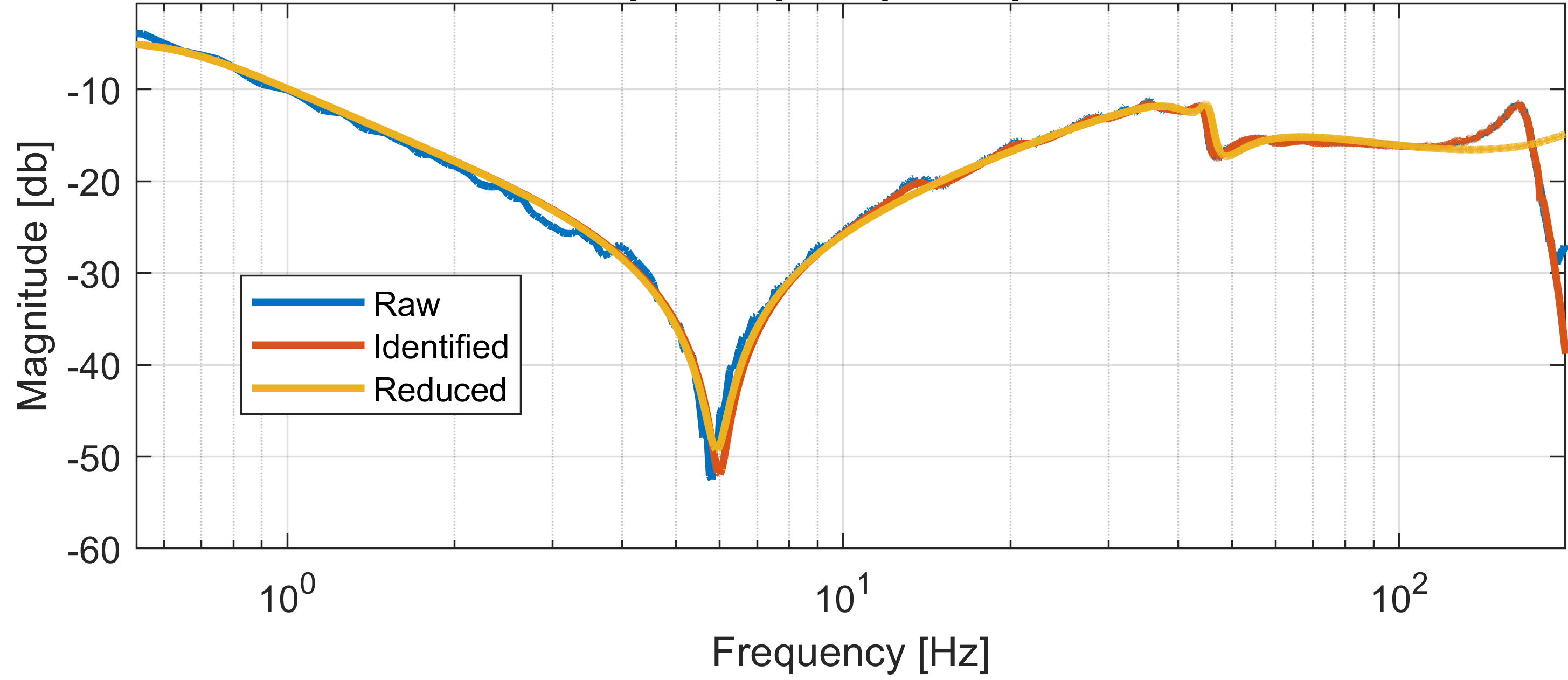}
   \end{tabular}
   \end{center}
   \caption[example] 
   { \label{fig:TFRed}
Open-loop transfer functions of the azimuth axis at ALT=50$^\circ$. Blue: raw noisy data; Red: 50-state identified model; Yellow: 14-state model after Hankel reduction.}
     \end{figure}

\section{System identification}
\label{sec:ID}  
The first step of the work has been the identification of the azimuth model. The derivation of the transfer function has been done purely from field test data, as no analytical model was available for comparison. A white noise signal has been applied at the torque input to excite the dynamics of the telescope. The white noise is a random signal of even spectrum, able to excite evenly each frequency component of the telescope dynamics. The corresponding velocity output has been measured by differentiating the position encoder output. 

It is worth noticing that, in the original implementation, the velocity feedback was provided by custom tachometers; after several years they started to fail and were replaced by an encoder based system, providing a back-compatible analogue speed signal. Finally, in the new system it has been decided to work with “software tachometers” obtained by differentiating the encoder readings. This is the same solution adopted, e.g. at the VLT (Ref.~\citenum{Erm2000}) and the VST (Ref.~\citenum{Sch12}).

To be effective, the white noise procedure has been tuned by varying parameters like the amplitude of the input noise and the duration of the test. After some tuning, the test was run at 500Hz frequency for an overall duration of 5 minutes, corresponding to 150,000 samples of input-output pairs. Longer acquisition times did not produce any difference. The amplitude of the input signal was determined after a trade-off between the opposite needs of exciting sufficiently the axis dynamics and prevent damages to the system. The average input signal was null, so the axis remained always approximately in the same position. 

The test configuration was extremely easy, as no external equipment was adopted. Everything was based on dedicated software features prepared in advance, fully integrated into the new graphical implementation of the control software, adopting a post-processing code developed in Matlab. The  white noise input was purely digital as well as the velocity output. The sample vectors were analyzed in quasi-real-time within the Matlab environment. Figure~\ref{fig:IdScheme} shows the basic diagram of the open-loop system identification. 

In the case of azimuth, it is well known that the transfer function depends on the altitude position (Ref.~\citenum{Gaw2008}). The model identification has been repeated at the two extremes of the altitude range (ALT=20$^\circ$ and ALT=90$^\circ$) and in an intermediate position (ALT=50$^\circ$). As expected, the three transfer functions are similar, but not identical (Figure~\ref{fig:TF}). The result at ALT=50$^\circ$ is assumed to be a good compromise, representing a first order approximation of the axis model. 

In principle, the identified model might be compared with an analytical model coming from FEA data, which would help to identify the physical meaning of all resonances in the model. Nevertheless, an accurate FEA model is not available for the TNG and any old simplified model, dating back to design time, could not be deemed reliable. Also, the purpose of this work is the design of the servo control for an existing telescope, rather than the study of a new optomechanical system still in design phase. Consequently, all the work was based uniquely on the identified transfer function. The identified model has been the starting point for the servo control design, described in the following sections. 

It is worth mentioning that for some control strategies (e.g. model-based control schemes) an accurate identified model is mandatory (Ref.~\citenum{Gaw2008}).  So far, the old servo control architecture has been kept for this work, but a model-based controller would be a viable alternative. Working on a model-based controller (e.g. LQG) design, it is generally necessary to reduce the order of the model to get an accurate, but simple, representation of the system. The order of the system must be decreased saving only an appropriate number of modes that still represent the system with good approximation. All the others, adding unuseful complexity to the controller design, shall be excluded. 

In the case under study, the azimuth axis was identified with a model of order 50, i.e. the dynamical matrix of the space-state representation is 50x50. This order is too high for the implementation of e.g. a LQG controller and likely includes dynamics that do not represent the system, possibly measurement noise. Different reduction techniques exist (Ref.~\citenum{Gaw04}); here the Hankel norm technique has been adopted, computing the Hankel singular vaues of the dynamical system and removing the modes with the smallest norms, under a given threshold. The threshold must be tuned in order to keep an adequate representation of the system. Figure~\ref{fig:TFRed} compares raw data, the 50-state azimuth model and a balanced 14-state  model reduced by truncation of the smallest Hankel norms; here the 50- and 14-state model differs significantly only for the high frequency dynamics, above 100Hz.

   \begin{figure} [ht]
   \begin{center}
   \begin{tabular}{c} 
   \includegraphics[height=15cm]{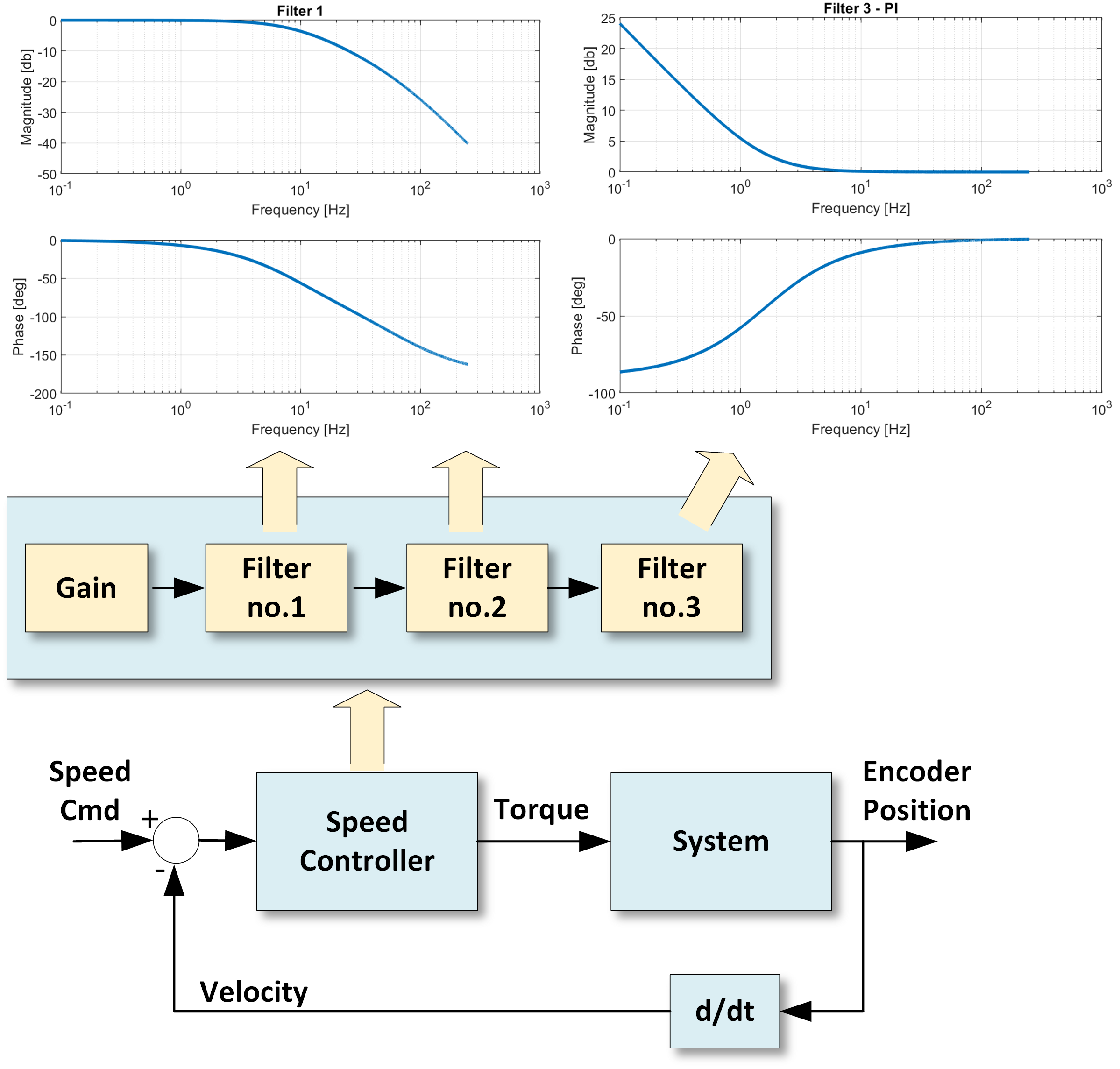}
   \end{tabular}
   \end{center}
   \caption[example] 
   { \label{fig:VelScheme}
The speed loop scheme. The inner structure of the speed controller is shown as a block diagram. On top, the Bode diagrams of the filters.}
     \end{figure}

   \begin{figure} [ht]
   \begin{center}
   \begin{tabular}{c} 
   \includegraphics[height=9cm]{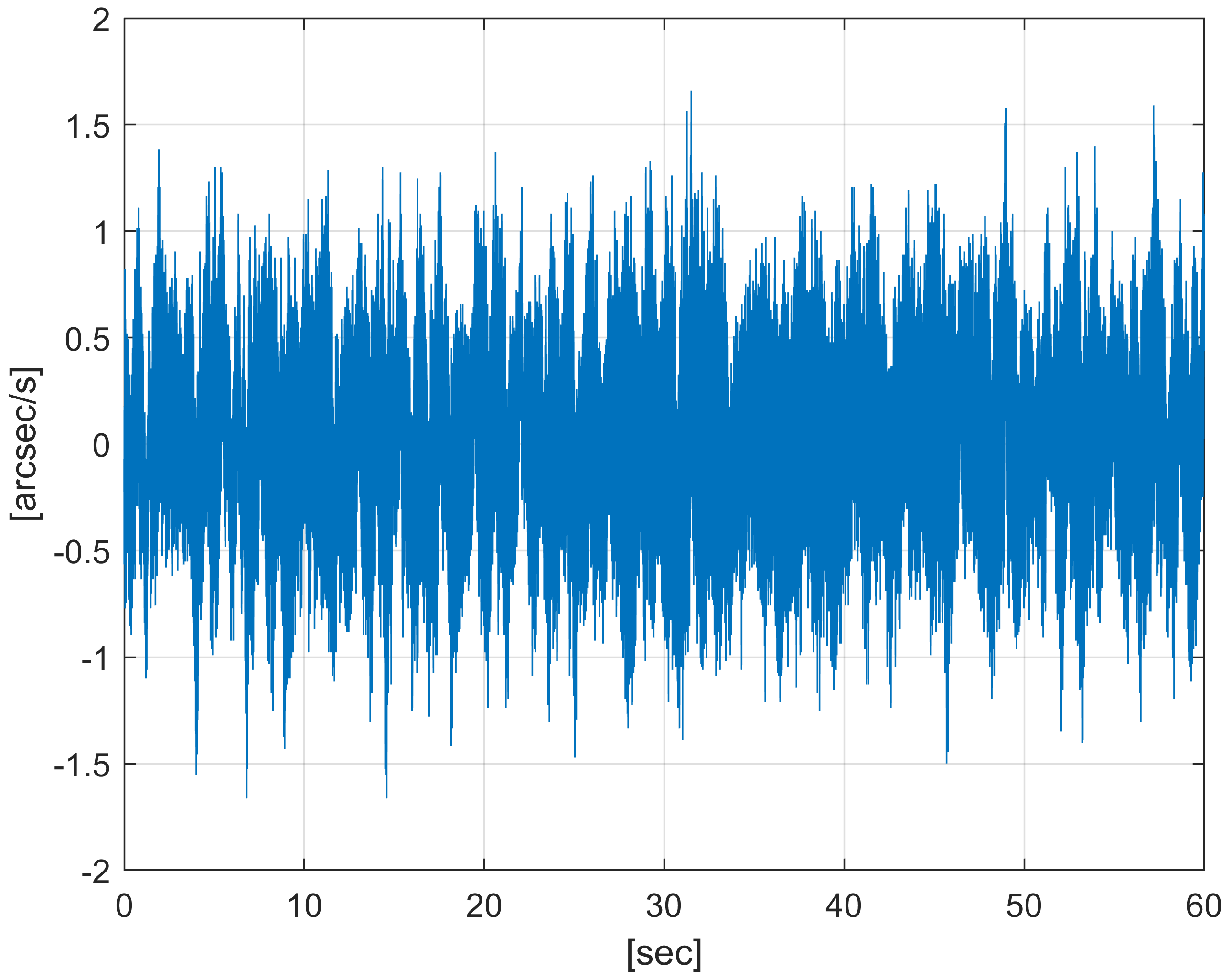}
   \end{tabular}
   \end{center}
   \caption[example] 
   { \label{fig:vErr}
Example of a velocity error time series, for a speed command of 80 arcsec/s (RMS=0.4 arcsec/s). }
     \end{figure}

\section{Speed Loop Control}
\label{sec:SL}  
The telescope main axes adopt a two-loop control structure with a speed loop nested inside the position loop. Adopting this structure, the problem is naturally decoupled in two steps, to be performed in the right order: 
\begin{itemize}
    \item design of the speed controller, which defines the closed-loop speed transfer function
    \item design of the position controller
 \end{itemize}  
 Thus, the speed controller design is the first step. A good speed control is necessary, but not sufficient, for a good star tracking. When the velocity control works sufficiently well, the control engineer can start dealing with the position control loop. 

The implementation of a series of second order filters offers a full flexibility for the speed controller design. They have an immediate interpretation, e.g. they can be low-pass, Notch, PID filters. The serial connection of a number of simple blocks allows to shape any possible transfer function in the frequency domain. Here, we preliminarily estimated that a series of four filters could give enough degrees of freedom. However, a good solution was identified using just three of them. The filters have been designed in the frequency domain, by placing poles and zeros where appropriate. In the current implementation, the filter no.1 and no.2 are identical low-pass filters and the filter no.3 is a PI controller. An external gain has been modulated in order to increase the bandwidth of the closed speed loop as much as possible, preserving simultaneously the system stability. Figure~\ref{fig:VelScheme} shows the speed closed-loop with a schematic description of the speed controller. Figure~\ref{fig:vErr} reports an example of speed error time series during the tuning phase.

\section{Position Loop Control}
\label{sec:PL}  
The fundamental goal for a telescope axis servo control is the quality of tracking. The control loops must be designed carefully to constrain the tracking error within tight specifications, especially for an optical telescope like the TNG. However, a good design of the slewing phase is essential as well. The slew phase must be as short as possible, so ideally the telescope axes should be driven at maximum speed, but this is impossible because of physical limits for the next position derivatives, i.e. acceleration and jerk. Overall, a system approach for servo control shall be adopted, addressing simultaneously all issues related to both slew and tracking phases. 

With reference to the PI (Proportional-Integral) controller, one of the most popular tracking position controllers for telescopes as well, one problem is represented by the large overshoot caused by the integral action during the slew phase, when the position error is large. One workaround is the implementation of different position controllers for slew and tracking, with a transition phase between the two; this was the first solution implemented at the VLT as described in Ref.~\citenum{Chiesa01} (later the implementation was changed as described in Ref.~\citenum{Sandrock12} adopting a command preprocessor, see also Section~\ref{sec:TG}).

\begin{figure} [ht]
   \begin{center}
   \begin{tabular}{c} 
   \includegraphics[height=4.5cm]{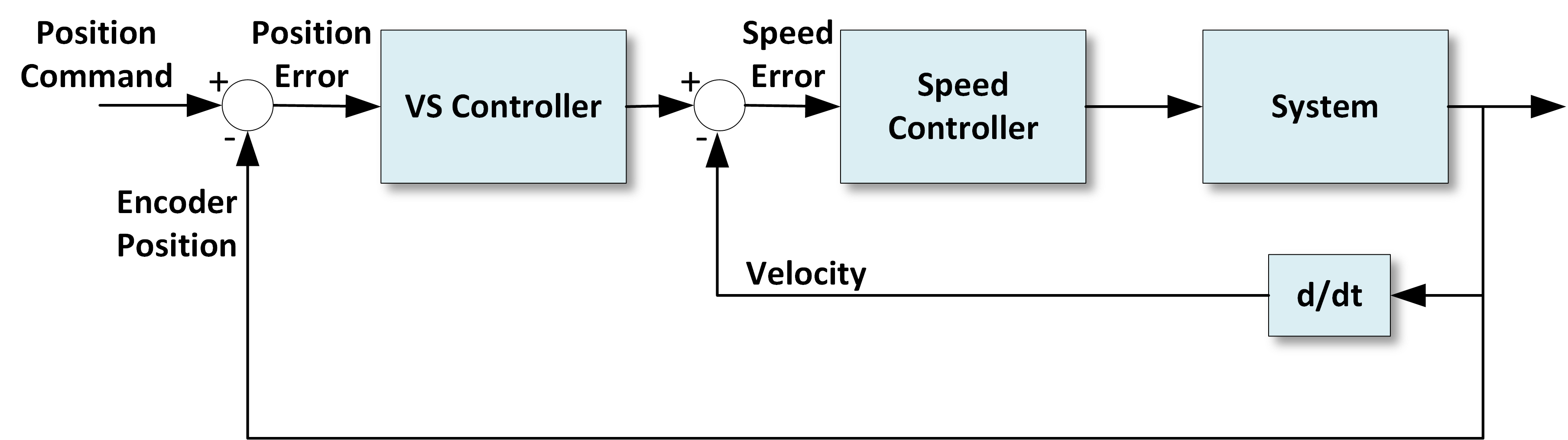}
   \end{tabular}
   \end{center}
   \caption[example] 
   { \label{fig:VSScheme}
Position loop adopting a Variable Structure controller.  }
     \end{figure}

A more sophisticated and flexible solution, still attractively easy, is the implementation of a variable structure controller in the loop (fig.~\ref{fig:VSScheme}), pioneered in the telescope field at the TNG itself in the nineties. With a variable structure controller, the parameters are dynamically changed as a function of the position error, with the following guidelines.
\begin{itemize}
    \item 
    When the axis is tracking and the error is small, the controller must guarantee optimal performance and minimal servo error. 
    \item
    When the error is large, using the tracking gains would cause instability. The position gain must be smaller and the integral gain must be zero for errors larger than a threshold.
    \item
     Between the two extreme conditions ({\em large} and {\em small} errors), it shall be possible to shape the gain functions as appropriate to allow for a smooth transition.
\end{itemize}

The implementation of these principles enables a smooth switch from a purely proportional to a proportional-integral controller, varying continuously the proportional and integral gains over the transition phase. 

\begin{figure} [ht]
   \begin{center}
   \begin{tabular}{c} 
   \includegraphics[height=8cm]{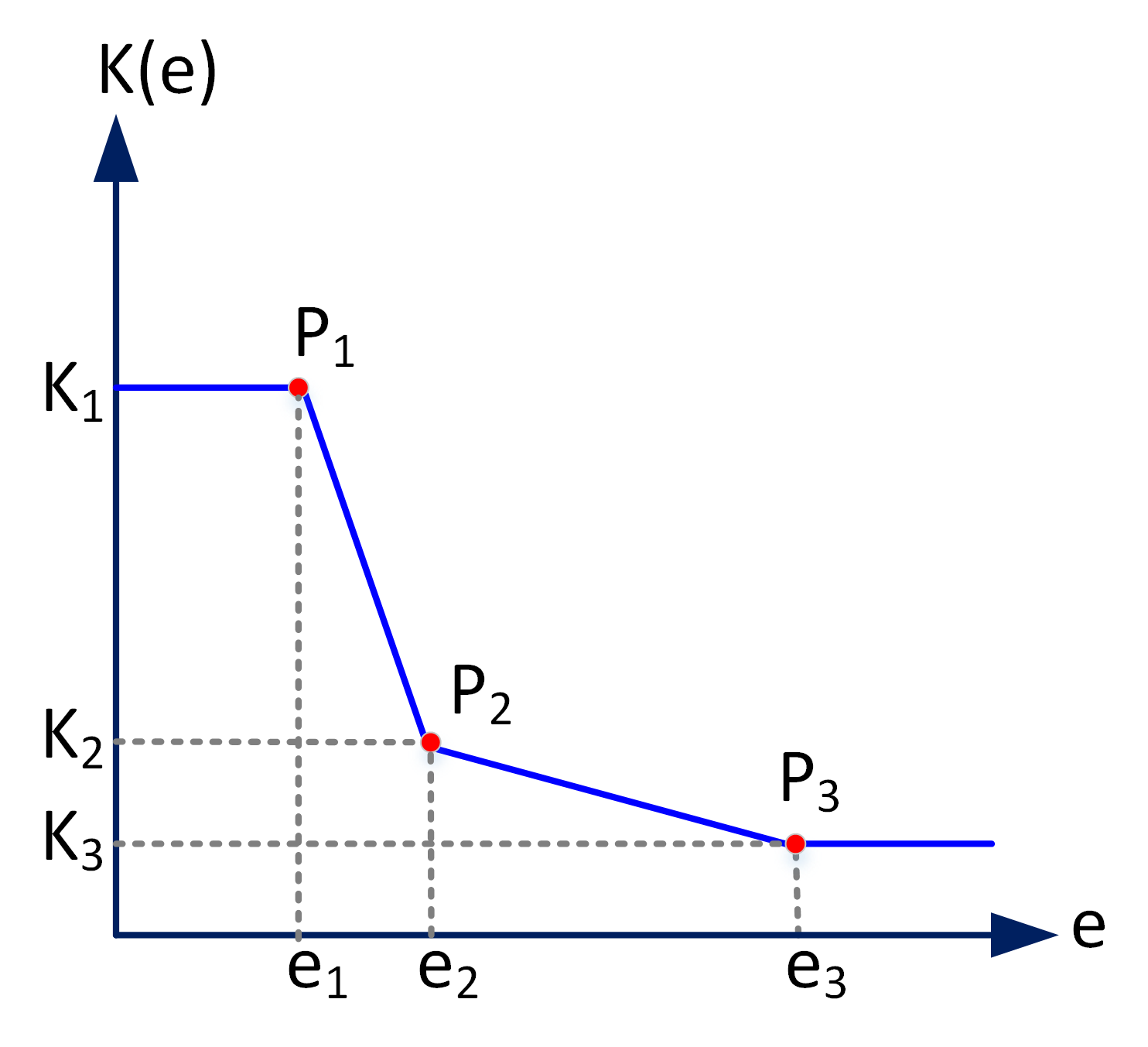}
   \end{tabular}
   \end{center}
   \caption[example] 
   { \label{fig:VS}
Variable structure controller gain. The K(e) function comes from the linear interpolation of  three points. The two extremes define the gain for the slewing phase (large errors) and the tracking phase (small errors); the central point gives the necessary degrees of freedom for the transition phase.}
     \end{figure}

In the TNG case, this has been implemented through a simple linear interpolation of three points in the gain-error plane; the principle is simply described by fig.~\ref{fig:VS}. The variable gain function is fully described by the coordinates of the three points, i.e. by 6 parameters. Thus, the PI variable structure controller is described by 12 parameters, fully defining Kp(e) and Ki(e). With reference to the transition phase, it is worth mentioning that more mathematically elegant connections of the two extreme points would not give, for the shaping of the functions, the same total freedom of a simple linear interpolation of points. In the TNG case, setting only one intermediate point has been enough.

   \begin{figure} [ht]
   \begin{center}
   \begin{tabular}{c} 
   \includegraphics[height=3.8cm]{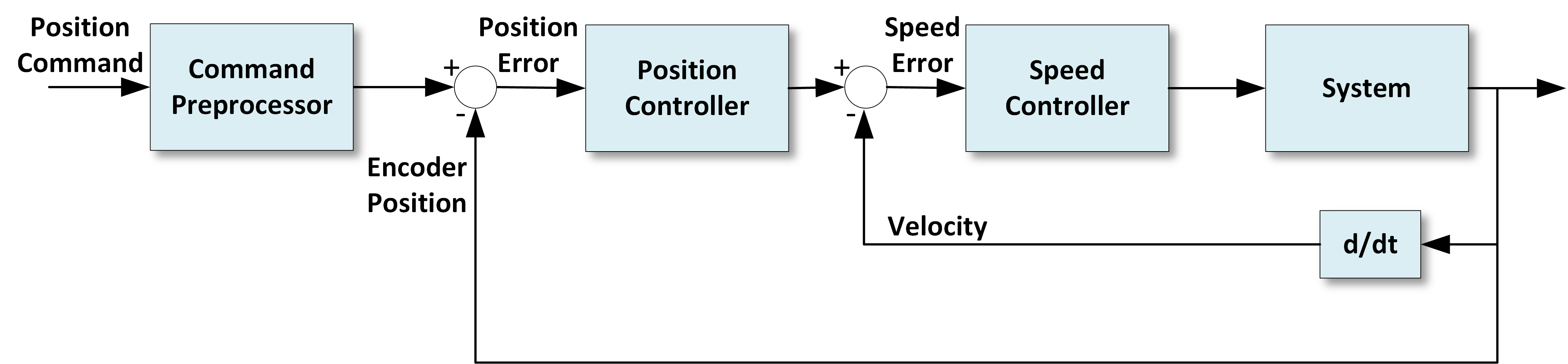}
   \end{tabular}
   \end{center}
   \caption[example] 
   { \label{fig:PrepScheme}
Position loop adopting a command preprocessor in input.  }
     \end{figure}

\section{Alternative implementations}
\label{sec:Fut}  

\subsection{Command preprocessor}
\label{sec:TG} 
Other approaches to the problems outlined in section~\ref{sec:PL} are possible, rather than recurring to variable structure controllers. An effective solution can be designing the slew phase by pre-filtering the axes reference positions coming from the  astrometric sequence, a solution sometimes referred to as {\em command pre-preprocessor} or {\em command shaper} (Fig.~\ref{fig:PrepScheme}). This way, the designer can constraint the system variables by planning a trajectory that will never exceed the velocity and acceleration limits, with two consequences: the telescope dynamics will always work in the linear regime, as any saturation will be prevented by design; the position error will always be limited, effectively preventing the overshoot problem caused by the integral action. 

The trajectory planning is not trivial because the telescope target is moving, although slowly, so the final destination of the telescope slew is not rigorously known in advance. Of course, in theory the whole trajectory of the celestial object might be precomputed; but this would require the servo control software embeds or replicates the astrometry computations. 

This is normally an impractical solution, because the two tasks of astrometry and servo control are usually decoupled and implemented separately: this is the case at the TNG too. An astrometry module computes the axes coordinates and generates the set-points for the servo control; on the servo side, the astrometry is seen as a black-box that outputs periodically the coordinates to track. Thus, the instantaneous trajectory set-points are known, but all the future ones are unknowns. Therefore, an essential feature for a command preprocessor at the TNG is the ability to work recursively, having no knowledge of the future positions but using instantaneous inputs only. 

Examples of slew trajectory planners are in Ref.~\citenum{Tyler98,Olberg95,Prestage97,Gaw02,Smith08, Hunter13, Savarese20}. They differ significantly in the approach and implementation. Here, the preference for a possible future implementation at the TNG goes for a recursive scheme guaranteeing the continuity of velocity and acceleration, i.e. for a preprocessor implementing a C${^2}$ class trajectory.

The implementation of a command shaper may lead to the implementation of a single position controller for any phase (large, medium and small errors), removing the need of variable gains.

\subsection{Model-based control}
The current PI controller improves the performance when the gains are increased, but makes the system unstable if the gains are too high, causing vibrations. The reason is that the telescope is not totally rigid but has flexible modes. A controller that  suppresses the telescope vibrations when the gains are increased would be helpful. This is what a model-based controller aims to do. Therefore, although the current implementation fulfills the requirements, moving to a model-based controller could in principle further enhance the system. 

One option is the LQG control. In the optimal control theory, a LQG controller is a combination of a Kalman filter (a Linear Quadratic state Estimator - LQE) with a Linear Quadratic Regulator (LQR).  The tracking version of the LQG controller described in Ref.~\citenum{Gaw2008} has some similarities with the PI control scheme and may be a candidate for an experimentation at the TNG.

\begin{figure} [ht]
   \begin{center}
   \begin{tabular}{c} 
   \includegraphics[height=5.7cm]{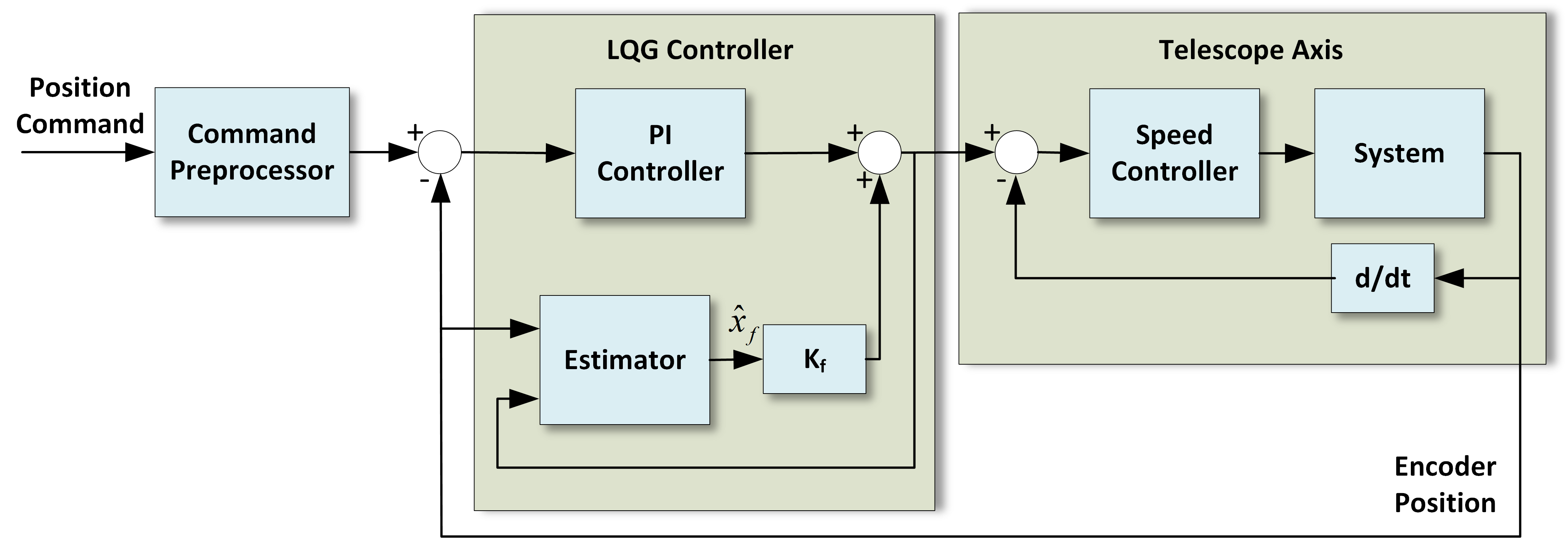}
   \end{tabular}
   \end{center}
   \caption[example] 
   { \label{fig:LQG}
Position loop adopting a command preprocessor and a LQG controller.  }
     \end{figure}

As shown in Fig.~\ref{fig:LQG}, the LQG is implemented in the position loop  and the "Telescope Axis", to be modelled through a white-noise identification procedure, is the velocity closed-loop. Here, the command preprocessor is mandatory: in fact the capability to shape the trajectory not exceeding velocity and acceleration limits allows to work always in the linear regime, as needed by this method. The scheme includes an estimator of the states of the flexible modes from the axis position and the speed command (output of position controller). The flexible mode controller $K_f$ action is added to the PI output, allowing the suppression of the telescope vibrations. 
The tracking version of the LQG controller is a PI augmented with a flexible mode controller, where the increase of the proportional gain cause no longer instability because the vibrations are suppressed. 

\section{Conclusions}
\label{sec:Conc}  
Within the framework of a hardware modernization effort at the TNG, more than two decades into the start of telescope operations, the servo control algorithms are consequently under update, to cope with the new drive system. After the control algorithms redesign and the validation of the tracking performance, the azimuth axis completed the transition to the new hardware. The open-loop axis model was identified and the servo controllers were redesigned from scratch. For the servo control algorithm redesign, few days of technical time in day-hours were allocated at the observatory, without stopping the telescope night operations. 

The minimum goal was leaving the telescope operational with the new hardware system, with at least the same satisfactory performance shown for the previous twenty years. This goal was achieved, as the final performance exceeded the previous ones. 

Given the limited time allocated for this first intervention, it was conservatively decided to start from the control philosophy implemented by the same authors at telescope commissioning time. New options are under study, which might further improve the TNG tracking quality.

\bibliography{tng} 
\bibliographystyle{spiebib} 

\end{document}